\documentclass[a4paper,longbibliography,twocolumn,superscriptaddress,prl, aps, floatfix,longbibliography ]{revtex4-1}

\usepackage[utf8]{inputenc}
\usepackage[english]{babel}
\usepackage{amssymb}
\usepackage{amsmath}
\usepackage{graphicx}
\usepackage{float}
\usepackage[normalem]{ulem}
\usepackage{color}

\newcommand{\eqcomma}{\,,}

\usepackage[percent]{overpic}

\begin{document}

\author{Kaj-Kolja Kleineberg}
\email{kkleineberg@ethz.ch}
\affiliation{
Computational Social Science, ETH Zurich, Clausiusstrasse 50, CH-8092 Zurich, Switzerland}
\author{Lubos Buzna}
\affiliation{
University of Zilina,
Univerzitna 8215/1, SK-01026 Zilina,
Slovakia}
\author{Fragkiskos Papadopoulos}
\affiliation{Department of Electrical Engineering, Computer Engineering and Informatics, Cyprus University of Technology, 33 Saripolou Street, 3036 Limassol, Cyprus}
\author{Mari\'an Bogu\~{n}\'a}
\affiliation{Departament de F\'isica de la Mat\`eria Condensada, Universitat de Barcelona, 
Mart\'i i Franqu\`es 1, 08028 Barcelona, Spain}
\affiliation{Universitat de Barcelona Institute of Complex Systems (UBICS), Universitat de Barcelona, Barcelona, Spain}
\author{M. Ángeles Serrano}
\affiliation{Departament de F\'isica de la Mat\`eria Condensada, Universitat de Barcelona, Mart\'i i Franqu\`es 1, 08028 Barcelona, Spain}
\affiliation{Universitat de Barcelona Institute of Complex Systems (UBICS), Universitat de Barcelona, Barcelona, Spain}
\affiliation{ICREA, Passeig Llu\'is Companys 23, E-08010 Barcelona, Spain}
\date{\today}

\title{Geometric correlations mitigate the extreme vulnerability of multiplex networks against targeted attacks}

\begin{abstract}
We show that real multiplex networks are unexpectedly robust against targeted attacks on high degree nodes, and that hidden interlayer geometric correlations predict this robustness. Without geometric correlations, multiplexes exhibit an abrupt breakdown of mutual connectivity, even with interlayer degree correlations. With geometric correlations, we instead observe a multistep cascading process leading into a continuous transition, which apparently becomes fully continuous in the thermodynamic limit. Our results are important for the design of efficient protection strategies and of robust interacting networks in many domains.
\end{abstract}

\maketitle

Networks are ubiquitous in many domains of science and engineering, ranging from ecology to economics, and often form critical infrastructures, like the Internet and financial systems. Nowadays, these systems are increasingly interdependent~\cite{helbing2013globally} and form so-called multiplex or multilayer networks~\cite{multilayer:kivel,ginestra:natphys}. This interdependency implies that, if a node fails in one network layer, its counterparts in the other layers also fail simultaneously. This process can continue back and forth between the layers, which makes them especially vulnerable to failures. In particular, an abrupt transition can arise in mutual percolation when nodes are removed at random~\cite{Buldyrev2010,ginestra:natphys,Gao2011}. Interestingly, interlayer degree correlations~\cite{pre:multiplex:correlations,pre:degree:correlations:2,prl:degree:correlations,scirep:degree:corr} mitigate this vulnerability to random node removals and the transition becomes continuous~\cite{Reis2014,self:similar:multiplex}.

In real systems, failures may not always be random but, instead, the result of targeted attacks. Multiplexes are extremely vulnerable to them on high-degree nodes~\cite{dong2013robustness,PhysRevE.89.042811,Zhao2016}, and exhibit a discontinuous phase transition even in the presence of interlayer degree correlations~\cite{PhysRevE.89.042811}.  
Although it is highly important for many real systems, it is not well understood how this vulnerability can be mitigated. 
Previous results point to negative interlayer degree correlations as a mitigation factor~\cite{PhysRevE.89.042811}, but real systems tend to show positive instead of negative interlayer degree correlations~\cite{pre:multiplex:correlations}. Are there other structural features that render multiplex networks robust against targeted attacks? And most importantly, are these properties present in real multiplexes?

Here, we show that interlayer hidden geometric correlations~\cite{geometry:multilayer} mitigate the vulnerability of multiplexes to targeted attacks. The removal of the highest degree nodes triggers multiple cascades which do not destroy the system completely, but eventually lead into a continuous percolation transition. Strikingly, we show that the strength of these geometric correlations in real systems is a good predictor of their robustness.

More specifically, we consider targeted attacks in two-layer multiplexes, where nodes are removed in decreasing order of their degrees among both layers. We rank all nodes $i$ according to $K_i = \max(k_i^{(1)},k_i^{(2)})$, where $k_i^{(j)}$ denotes the degree of node $i$ in layer $j=1, 2$. 
We remove nodes with higher $K_i$ first (we undo ties at random) and re-evaluate all $K_i$s after each removal. To measure the percolation state of the multiplex, we compute its mutually connected component (MCC) as the largest fraction of nodes that are connected by a path in every layer using only nodes in the component~\cite{Buldyrev2010}. 

\begin{figure}[t]
\begin{overpic}[width=0.47\textwidth]{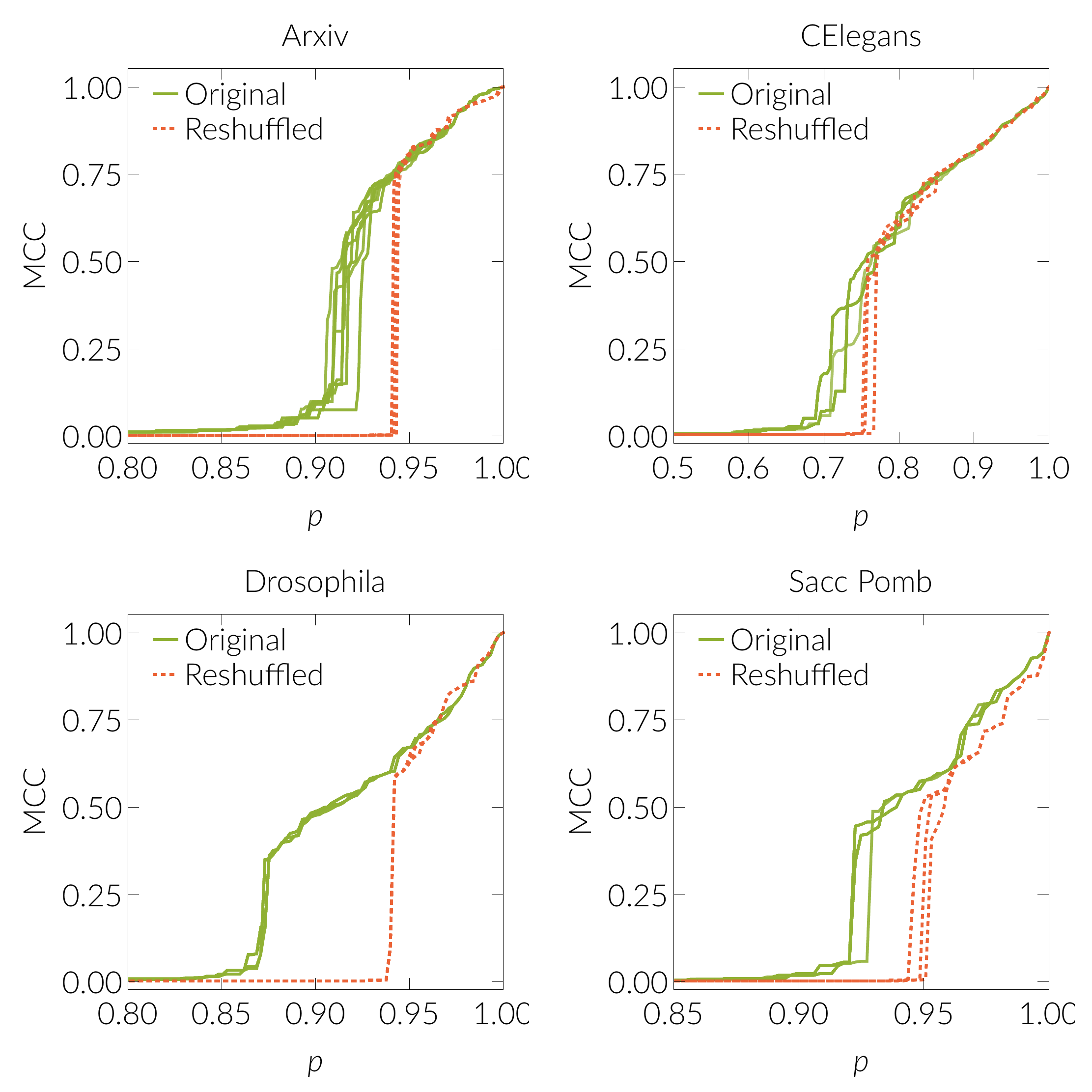}
 \put (0,95) {\large \textsf{a)}}
  \put (50,95) {\large \textsf{b})}
   \put (0,45) {\large \textsf{c)}}
  \put (50,45) {\large \textsf{d})}
\end{overpic}
 \caption{ \textbf{(a)}
 {\color{black}Relative size of the mutually connected component (MCC) against the fraction $p$ of nodes remaining in the system}
 for the arXiv (layers 1, 2) collaboration multiplex (green lines) and for its reshuffled counterpart (red dashed lines). Different lines correspond to different realizations of the targeted attacks process.
 \textbf{(b)} shows the same for the C. Elegans multiplex (layers 2 and 3),
 \textbf{(c)} for Drosophila (layers 1 and 2), and 
 \textbf{(d)} for Sacc Pomb (layers 3 and 4).
 \label{fig_arx_and_ce}
 }
\end{figure}

\begin{figure*}
\includegraphics[width=1\linewidth]{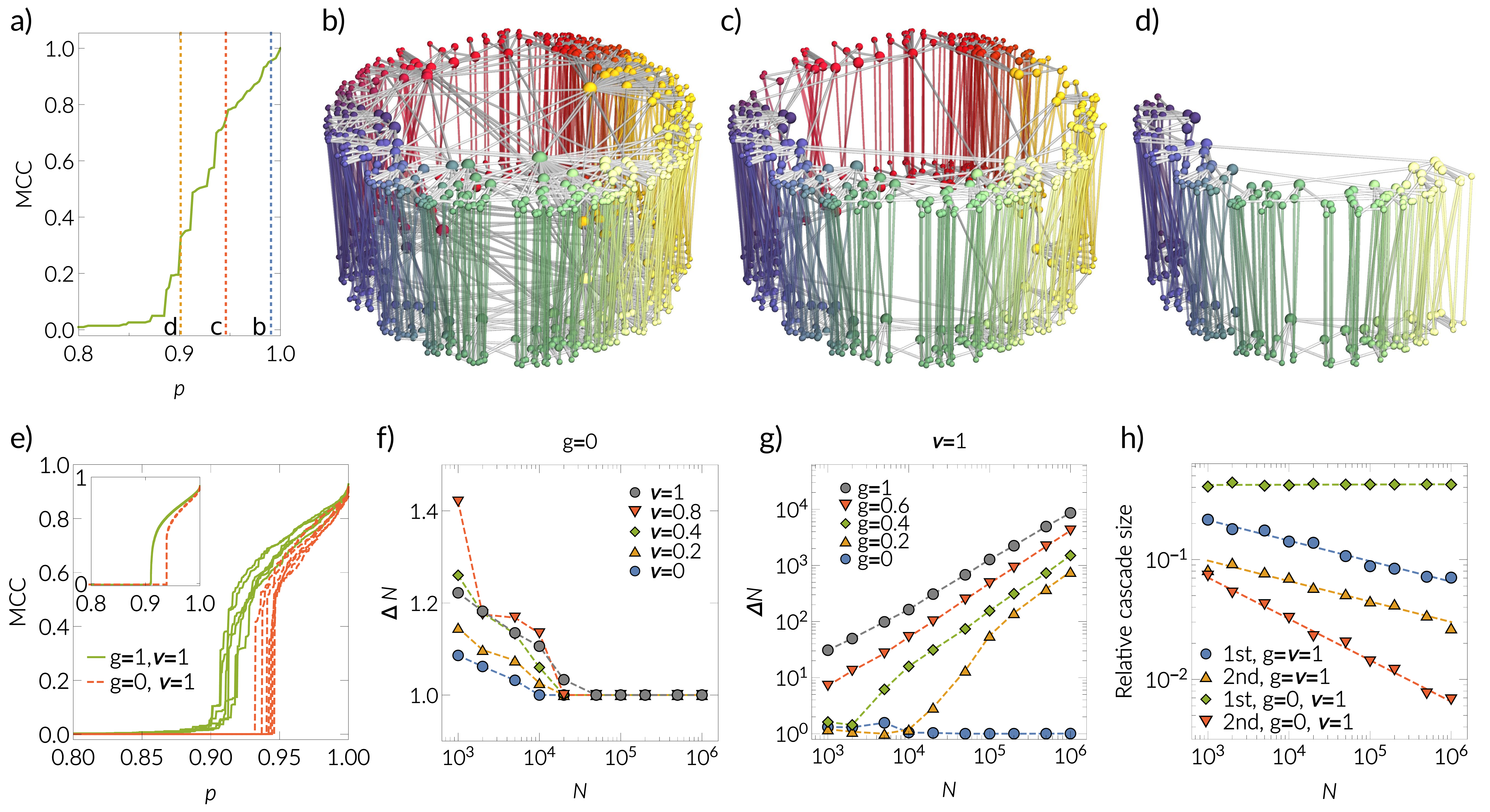}
\caption{Targeted attacks on synthetic multiplex networks generated by the GMM model (see text). Each layer has power law degree distribution with exponent $\gamma = 2.6$, average node degree $\left< k \right> \approx 6$, and clustering $\bar{c} = 0.35$.
\textbf{(a-d)}
Here, each layer has $N=500$ nodes, and we have set $g=1$ and $\nu=0$. 
\textbf{(a)} relative size of the MCC as a function of the fraction of nodes remaining in the system $p$. 
\textbf{(b)} MCC after the removal of $4$ nodes (corresponding to the dashed blue line in (a)). 
\textbf{(c)} the same as in (b) after the removal of $23$ nodes (dashed red line in (a)). 
\textbf{(d)} the MCC after the removal of $42$ nodes  (dashed yellow line in (a)).
\textbf{(e)} Evolution of the MCC in a two-layer synthetic multiplex with layers of size $N=2 \times 10^3$ nodes.
The inset shows the same for $10^6$ nodes.
\textbf{(f)} critical number of nodes, $\Delta N$, as a function of the system size $N$ when there are no angular correlations, $g=0$, and for different radial correlation strengths. The results are averages over $60$ realizations {\color{black}(for $N<10000$ we performed $1000$ realizations)}. 
\textbf{(g)} the same as in (f) but for different values of the angular correlations strength $g$ and for fixed $\nu=1$.
\textbf{(h)} shows the largest and second largest cascade size (relative to system size). 
\label{fig_il}}
\end{figure*}
\begin{figure}[t]
\includegraphics[width=0.97\linewidth]{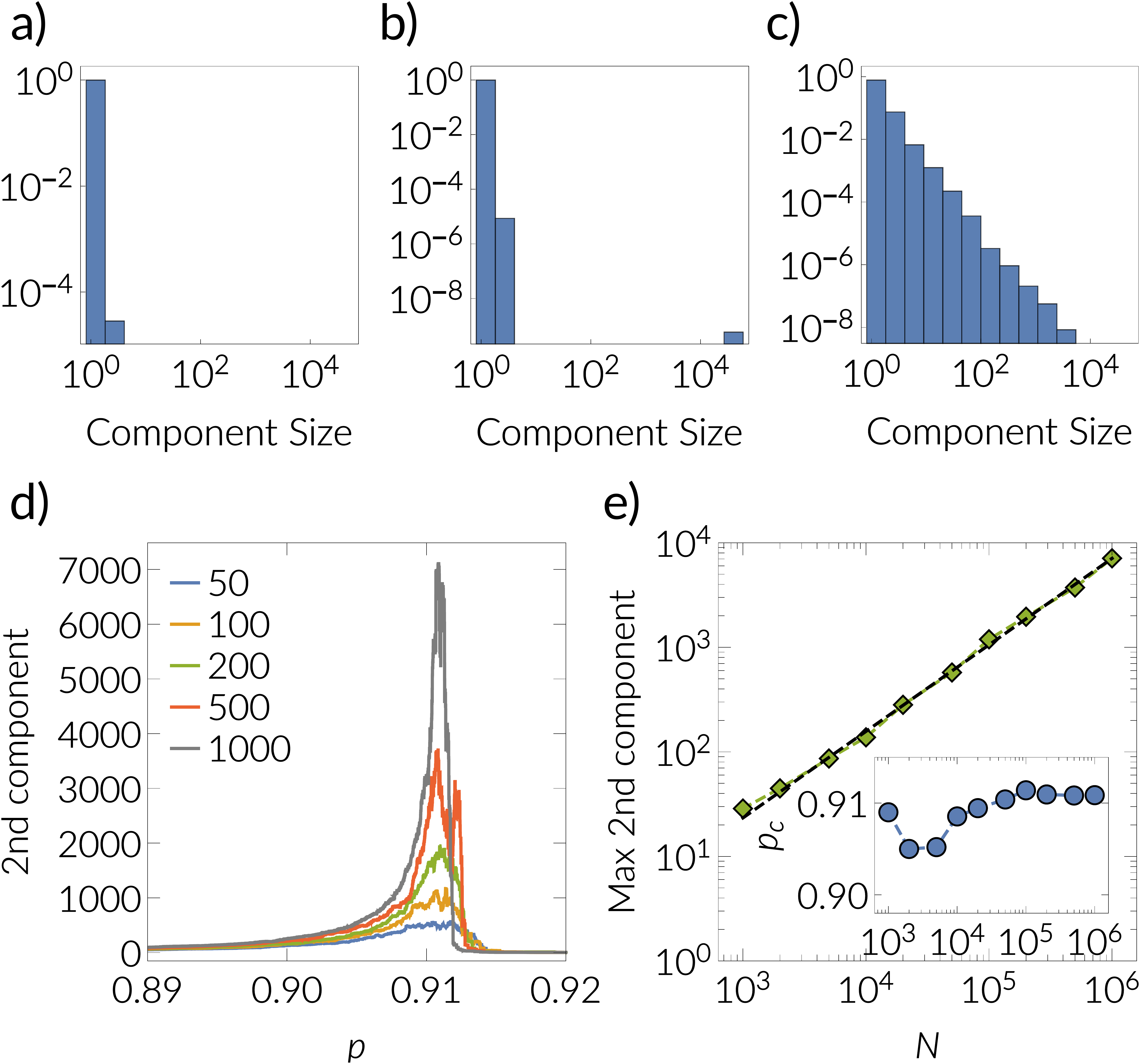}
\caption{
{\color{black} \textbf{(a-c)} shows the distribution of component sizes (PDF) during the evolution of the MCC for two-layer synthetic multiplexes constructed with the GMM model. Each layer has a power law degree distribution with exponent $\gamma=2.6$, average node degree $\left < k \right> = 6$, and clustering $\bar{c} = 0.35$. In \textbf{(a-c)} each layer has $N=5 \times 10^4$ nodes.
\textbf{(a)} distribution of component sizes directly before the transition ($p = 0.94539$), and \textbf{(b)} directly after ($p = 0.94540$), when there are no radial or angular correlations, $\nu=0, g=0$.
\textbf{(c)} distribution of component sizes at $p = 0.9078$ when there are maximal angular correlations, $g=1$, and no radial correlations, $\nu=0$.
\textbf{(d)} absolute size of the second largest MCC as a function of $p$ for different layer sizes $N$ as indicated in the legend ($\times 10^3$); for each size, the results are averages over $60$ realizations of the multiplex (as in \textbf{(e)}).
\textbf{(e)} scaling of the maximum of the second largest MCC. The black dashed line shows a fit $\propto N^{0.84}$, while the inset shows the value of $p = p_c$ where the maximum is realized.
}
\label{fig_model}}
\end{figure}

Figure~\ref{fig_arx_and_ce} shows results
for the real arXiv collaboration~\cite{prx:modular}, C. Elegans~\cite{muxviz}, Drosophila~\cite{arenas:reduce}, and Sacc Pomb~\cite{arenas:reduce}
(see Table~\ref{tab_data}, SM Section I, and Supplementary Videos I-IV) as well as for their reshuffled counterparts (an illustration of a targeted attack sequence is shown in Fig.~\ref{fig_il}a-d). 
To create the reshuffled counterpart, we randomly reshuffled the translayer node-to-node mappings by selecting one of the layers and randomly interchanging the internal IDs of the nodes in that layer. This process destroys all correlations between the layers without altering the layers' topologies~{\color{black}(see SM Section~I for further details)}. 
\begin{table}[t]
 \begin{ruledtabular}
\begin{tabular}{lllll}
  \text{Dataset} & \text{MCC} & $\Delta N$ & $\Delta N_{rs}$ & NMI \\ \hline
 \text{arXiv Layers 1, 2} & 790 & 25.2 & 1.0 & 0.58 \\
 \text{Physicians Layers 1, 2} & 104 & 6.0 & 1.0 & 0.41 \\
 \text{Internet Layer 1, 2} & 4710 & 81.4 & 14.1 & 0.34 \\
 \text{C. Elegans Layers 2, 3} & 257 & 14.0 & 1.1 & 0.34 \\
 \text{SacchPomb Layers 3, 4}  & 426 & 4.2 & 1.5 & 0.17 \\
 \text{Drosophila Layers 1, 2} & 449 & 8.4 & 2.0 & 0.26 \\
 \text{Brain Layers 1, 2} & 74 & 7.0 & 1.0 & 0.19 \\
 \text{Rattus Layers 1, 2} & 158 & 4.0 & 1.0 & 0.18 \\
 \text{Air/Train Layers 1, 2} & 67 & 3.0 & 3.0 & 0.10 \\
\end{tabular}
 \end{ruledtabular}
 \caption{ 
Analyzed datasets for selected layer pairs (see SM Section~I for all layer pairs). 
MCC denotes the initial size of the MCC, $\Delta N$ denotes the critical number of nodes whose removal reduces the MCC from 
$40\%$ to $\sqrt{M}/M$ (in relative size),
and $\Delta N_{rs}$ the same for the reshuffled system. Values are averages over $100$ realizations of the removal process. NMI denotes the normalized mutual information {\color{black} (see SM Section IX)}
and gives a measure of the strength of angular correlations between the layers of the considered real systems.
 \label{tab_data}}
\end{table}
We quantify the \emph{vulnerability} of the real and reshuffled multiplexes by calculating the critical number of nodes, $\Delta N$.
The removal of this critical number reduces the size of the MCC from more than $a M$ 
to less than $M^{\beta}$, where $M$ is the initial size of the MCC before any nodes are removed, {\color{black} $a \leq 1$} is a threshold parameter, and $\beta < 1$~\cite{Achlioptas2009}. We set $a=0.4, \beta=0.5$. The larger the $\Delta N$, the more robust (less vulnerable) the system is. 
For the real arXiv multiplex we find that $\Delta N \approx 25$, while for its reshuffled counterpart $\Delta N_{rs} = 1$. In fact, in the reshuffled system, the removal of a single node reduces the relative size of the 
MCC from $73\%$ to only $0.25\%$. This is far more pronounced than the limits of {\color{black}$a=40\%$} and $\sqrt{M}/M=3.6\%$,
and is enough to virtually disconnect this system. 
We have considered other layer pairs of the arXiv, as well as a large number of other real multiplexes from different domains (technological, social, and biological). We found that in the vast majority of cases, the real system is significantly more robust against targeted attacks than its reshuffled counterpart (see Table~\ref{tab_data} and SM Sections I, II).

Below, we show that this increased robustness of real multiplexes to targeted attacks is due to hidden geometric correlations interwoven in their layers~\cite{geometry:multilayer}, which do not exist in their reshuffled counterparts. 
{\color{black} Specifically, each single network layer can be mapped (or embedded) into a separate hyperbolic space~\cite{Boguna2010, frag:hypermap, frag:hypermap_cn}, where each node $i$ is represented by its radial (popularity) and angular (similarity) coordinates, $r_i, \theta_i$, which are both significantly correlated in different layers, while hyperbolically closer nodes in each layer are connected with higher probability (see SM Section~I for further details).} 

Radial correlations are equivalent to interlayer degree correlations~\cite{Krioukov2010}. Angular correlations, instead, lead to sets of nodes that are similar---close in the angular similarity space---in each layer of the multiplex~\cite{geometry:multilayer}. The reshuffling process explained earlier destroys both radial and angular correlations between the layers. The extreme vulnerability of the reshuffled counterparts in comparison to the real systems raises fundamental questions: Are the radial (i.e., interlayer degree) correlations, or angular (i.e., geometric) correlations, or both, responsible for the robustness of real systems, and which of these correlations can help to avoid catastrophic cascading failure when multiplexes are under targeted attack? 

To investigate these questions, we use the geometric multiplex model (GMM) {\color{black}(see SM Section~III)}
to generate synthetic two-layer multiplexes, which resemble the real equivalents. The model produces multiplexes with layers embedded into hyperbolic planes, whereby the strength of interlayer correlations between the radial and angular coordinates of nodes that simultaneously exist in both layers can be tuned by varying the model parameters $\nu \in [0,1]$ and $g \in [0,1]$. Radial correlations increase with parameter $\nu$ ($\nu = 0$ for no radial correlations, whereas $\nu = 1$ for maximal radial correlations). Similarly, angular correlations increase with parameter $g$ ($g = 0$ for no angular correlations, while $g = 1$ for maximal angular correlations). 

We find that synthetic multiplexes without angular correlations exhibit an extreme vulnerability to targeted attacks (see Fig.~\ref{fig_il}e, SM Section III, and Supplementary Video V), similarly to the reshuffled counterparts of real systems (cf. Fig.~\ref{fig_arx_and_ce} and SM Section II).  
In particular, if the multiplex is sufficiently large, then the removal of only a single node can reduce the size of the MCC from $40\%$ to the square root of its initial size, thus destroying the connectivity of the system, see Fig.~\ref{fig_il}f. 
The abrupt character of the transition is also reflected in the distribution of mutually connected component sizes. In the fragmented phase,
the entire network is always split into very small components, even when the system is very close to the transition ({\color{black} see Fig.~\ref{fig_model}a and SM Section~IV}).
In the percolated phase, only nodes that do not belong to the MCC remain fragmented into small components ({\color{black} see Fig.~\ref{fig_model}b and SM Section~IV)}. This behavior
is not affected by the strength of the radial (i.e., interlayer degree) correlations in the system. Thus, in contrast to the mitigation effect for random failures, interlayer degree correlations do not avoid an abrupt transition in the case of targeted attacks, and essentially do not affect the robustness of the system. 

On the other hand, this extreme vulnerability is mitigated if angular correlations are present. In Fig.~\ref{fig_il}a-d and e,
we show the MCC percolation transition for maximal angular correlations (see also SM Sections II, III, and Supplementary Video V). 
We observe that the transition does indeed start with a multistep cascading process for relatively small system sizes. 
{\color{black}
However, as shown in Fig.~\ref{fig_il}f and Fig.~\ref{fig_il}g, the critical number of nodes, $\Delta N$, scales with the system size in the presence of angular correlations, see also SM Section V, while it always converges to one for large system sizes if angular correlations are absent. Moreover, as shown in Fig.~\ref{fig_il}h, the relative size of the largest jump after a single node removal decreases with the system size, in stark contrast to the case without angular correlations, where this quantity becomes size independent. This suggests that, in the thermodynamic limit, the system undergoes a continuous transition (see inset in Fig.~\ref{fig_il}e).
}
Furthermore, the size of the second largest component scales with the system size like $N^\sigma$, with $\sigma \approx 0.84$ ({\color{black} see Fig.~\ref{fig_model}d,~e and SM Section~VI)}. Finally, at the transition, the distribution of component sizes follows a power-law ({\color{black} see Fig.~\ref{fig_model}c and SM Section~IV}).
Thus, we conjecture that angular correlations can lead to a multistep cascading process for relatively small system sizes, and can give rise to a continuous transition in the thermodynamic limit (happening in a range of parameters of the model---including those used in {\color{black}Fig.~\ref{fig_il}}---such that the multiplex layers have strong metric structure but do not loose the small-world property in the targeted attack process, see SM Section~VII). 
This behavior is not affected by the strength of radial (i.e., interlayer degree) correlations and cannot be explained by the link overlap induced by geometric correlations (see SM Section VIII).
Taken together, our results suggest that angular (similarity) correlations can mitigate the extreme vulnerability of real multiplexes against targeted attacks.

\begin{figure}[t]
  \includegraphics[width=1\linewidth]{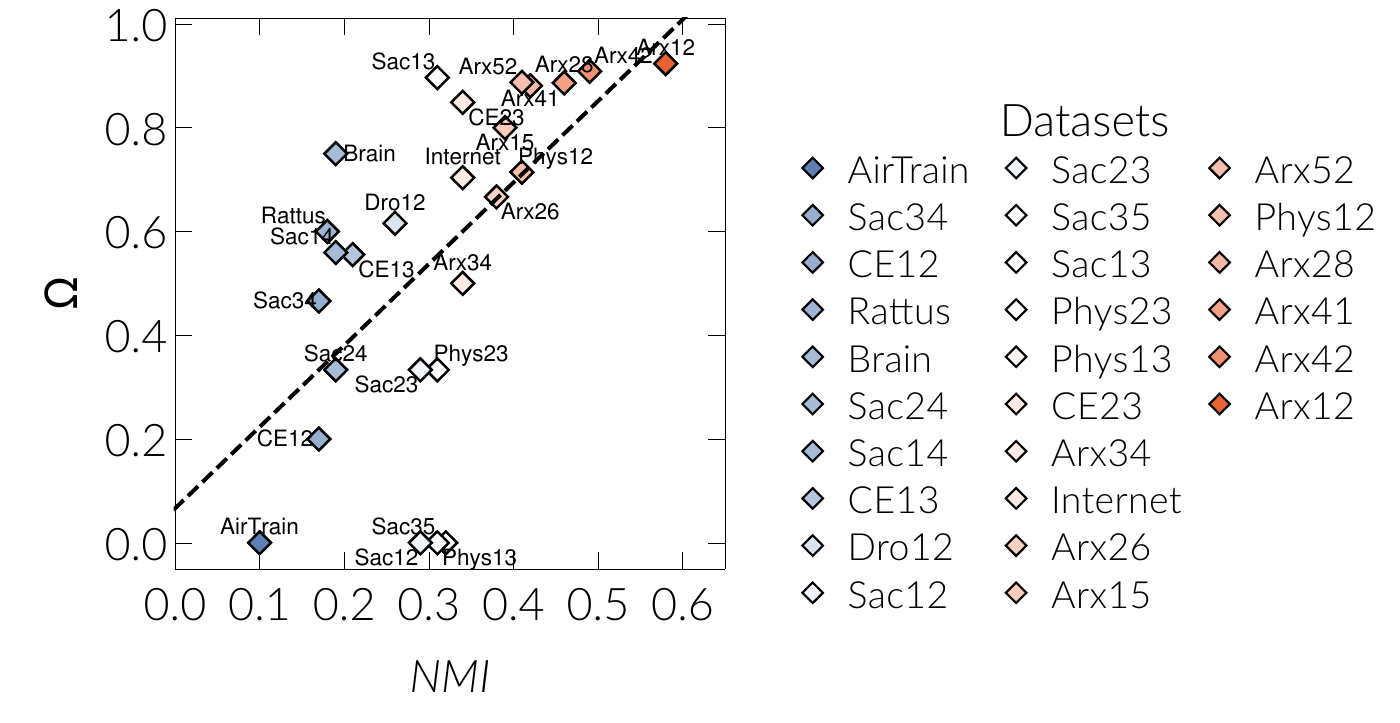}
 \caption{Relative mitigation of vulnerability $\Omega$ (Eq.~(\ref{eqn_omega})) as a function of the normalized mutual information $NMI$, which is a measure of the strength of angular correlations between the layers of the considered real systems (see SM Section IX for details). 
 \label{fig_nmi_vs_omega}
 }
\end{figure}

We can validate this conclusion in real systems. To this end, we compare the vulnerability of each of the considered real multiplexes (see Table~\ref{tab_data} and SM Section I) with that of its reshuffled counterpart. We define the relative mitigation of vulnerability as
\begin{equation}
\Omega = \frac{\Delta N - \Delta N_{\text{rs}}}{\Delta N + \Delta N_{\text{rs}}} \eqcomma
\label{eqn_omega}
\end{equation}
where $\Delta N$ and $\Delta N_{\text{rs}}$ are the number of nodes needed for the critical reduction of the size of the MCC of the real and reshuffled systems, see Table~\ref{tab_data} and SM Section~I. $\Omega$ is a measure of how much more resilient the real networks are compared to their reshuffled counterparts. Next, we study how $\Omega$ behaves as a function of the strength of angular correlations in the considered real systems. 
We quantify the strength of interlayer angular correlations by calculating the normalized mutual information, $NMI$, between the inferred angular coordinates of nodes in different layers
{\color{black}(see SM Section IX)}. A larger $NMI$ means higher angular correlations.
We find a strong positive correlation ($\rho \approx 0.6$)
between the strength of angular correlations in the real systems and their relative mitigation of vulnerability, see Fig.~\ref{fig_nmi_vs_omega}. This finding validates our previous arguments with real data, and highlights the importance of angular correlations in making real multiplexes robust against targeted attacks.

The gain of robustness due to angular correlations can be understood intuitively by the formation of macroscopic mutually connected structures on the periphery of the hyperbolic disc in each layer.
After enough nodes are removed, the remaining multiplex resembles a ``double ring'' (Fig.~\ref{fig_il}c), because the higher degree nodes which have been removed had lower radial coordinates and hence were closer to the center of the disc. If angular correlations are present, the remaining lower degree nodes that are close in one layer tend to also be close in the other layer. As a consequence, the double ring contains macroscopic mutually connected structures (Fig.~\ref{fig_il}d) that sustain connectivity in the system. 
Notice that the mitigation of the extreme vulnerability of multiplexes by the effect of angular correlations is directly related to their geometric nature and cannot be explained by any topological feature. To support this point, we checked whether interlayer clustering correlations (being clustering the topological feature which is more directly related to the metric properties of networks~\cite{Serrano2008}) or edge overlap induced by geometric correlations are sufficient to produce the mitigation effect. The results, see SM Sections VIII and X, indicate that in the absence of angular correlations, neither clustering correlations nor overlap can explain the observed mitigation effect.
We take this to be a new validation of the geometric nature of complex networks and of the role of geometric correlations in multiplexes.
 
To conclude, we have shown that the strength of geometric (similarity) correlations in real multiplex networks is a good predictor for their robustness against targeted attacks, providing, for the first time, strong empirical evidence for the relevance of this mechanism in real systems. Using a geometric multiplex network model, we have shown that multiplex networks are extremely vulnerable against targeted attacks, exhibiting a discontinuous phase transition if geometric (similarity) correlations are absent. Contrarily, the presence of such correlations mitigates this vulnerability significantly, inducing a multistep cascading process in relatively small systems which does not destroy the system completely but lead into an eventually smooth percolation transition, with results suggesting that it can be fully continuous in the thermodynamic limit. 
In particular, the critical number of nodes that has to be removed to disconnect the system scales with the system size only if geometric correlations are present. 
Our results can help when designing efficient protection strategies and more robust and controllable interdependent systems. In addition, the results highlight that dependent networks without similarity correlations are extremely vulnerable to targeted attacks.
Finally, our findings pave the way for an exact analysis of the percolation properties of such systems via their hidden geometric spaces.

\begin{acknowledgments}
K-K. K. acknowledges support by the ERC Grant ``Momentum'' (324247); 
L. B. has been supported by projects VEGA 1/0463/16, APVV-15-0179 and FP 7 project ERAdiate (621386);
F. P. acknowledges support by the EU H2020 NOTRE project (grant 692058).
M.~A.~S. and M.~B. acknowledge support from a James S. McDonnell Foundation Scholar Award in Complex Systems, MINECO projects no.~FIS2013-47282-C2-1-P and no.~FIS2016-76830-C2-2-P (AEI/FEDER, UE), and the Generalitat de Catalunya grant no.~2014SGR608. M.~B. acknowledges support from the ICREA Academia prize, funded by the Generalitat de Catalunya.
\end{acknowledgments}

\end{document}